# Identification of logical errors through Monte Carlo simulation

Hilary L. Emmett, Decisioneering (UK) Ltd.
Broadgate Court, 199 Bishopsgate, London EC2M 3TY, United Kingdom
hemmett@crystalball.com

Lawrence I. Goldman, Decisioneering Inc.
1515 Arapahoe Street, Suite 1311, Denver, CO 80202, U.S.A.
lgoldman@crystalball.com

**ABSTRACT**

*The primary focus of Monte Carlo simulation is to identify and quantify risk related to uncertainty and variability in spreadsheet model inputs. The stress of Monte Carlo simulation often reveals logical errors in the underlying spreadsheet that might be overlooked during day-to-day use or traditional "what if" testing. This secondary benefit of simulation requires a trained eye to recognize warning signs of poor model construction.*

## 1 INTRODUCTION

### 1.1 Types of Spreadsheet Risk

Risk is commonly defined as the probability of loss, damage, or any other undesirable event. For spreadsheet users, risk is the probability of the spreadsheet model yielding an incorrect out put. Here, we will divide the sources of risk into two major categories, those due to incorrect inputs, and those due to faults in the mathematical and logical formulas within the model.

When using spreadsheets, analysts traditionally input average or best-guess values for uncertain variables because programs like Microsoft® Excel only allow them to enter a single value or formula in a cell. These "deterministic," or known, models provide a single outcome upon which a business or technical decision is made. To capture uncertainty, analysts can perform simple "what if analysis" or "scenario analysis" by manually changing model variables and analyzing their effect on the key outputs. This approach provides a range of possible outcomes but does not impart an understanding of the likelihood of any particular outcome. There are simply too many combinations of input values to calculate every possible result.

### 1.2 Monte Carlo Simulation

One established solution to the limitations of spreadsheet risk analysis is Monte Carlo simulation. Since Excel alone does not have the ability to run and analyze simulations, modelers must rely on third-party programs like Crystal Ball® that add in and expand the features of Excel. Crystal Ball adds two techniques to Excel: the replacement of single values with probability distributions and the random simulation of a model. The result is a probability-based spreadsheet with quantifiable outcomes, such as a 75% probability of staying under budget or a 90% certainty of 100 million barrels of oil within a geologic reservoir. This example focuses on the probability of success for a simple project, based on the uncertainty in future sales, cost of goods sold, and operating expenses. Excel's NPV function is used to calculate the present value of the project.

As an auditing tool, Monte carlo simulation tests data outside the limits of the normal range (distribution tails.) In a survey of Australian spreadsheet developers, 63% acknowledged this as something they should be doing, while only 33% actually reported using the technique. [Hall, 1996]

**1.3 Garbage In, Garbage Out**

One of the major criticisms of Monte Carlo simulation is that it can provide a false sense of confidence in cases where the underlying spreadsheet logic is incorrect. This problem is prevalent in Excel-based tools, where the stochastic results are only as good as the skill, accuracy, and discipline of the spreadsheet developer.

There are, however, several check points where the output of a simulation can reveal formula errors that had been overlooked in the process of creating the spreadsheet. These check points occur during and after the simulation and through the analysis of several different types of charts and statistical methods.

Following recent accounting scandals in the United States, the Securities and Exchange Commission has mandated some form of risk analysis for all publicly traded companies. Similarly, the Financial Accounting Standards Board now requires reporting on the value of employee stock options. A number of companies are now implementing Monte Carlo simulation in an attempt to comply with these standards.

In comparison, formal spreadsheet application review and audit mandates are estimated to exist in as few as 10% of companies. [Hall, 1996] Any attempt to formalize the spreadsheet development process can be seen as an improvement. Model verification through simulation is not intended as a substitute for a structured development and review of all model logic, but rather as a final reality check.

**2 TOOLS FOR RELATING RISK INTO ACTION**

**2.1 Modeling Input Risk**

The first step in the move from deterministic to stochastic modeling is to identify key input variables that are subject to uncertainty. We will refer to these inputs as assumptions. For each assumption, the user must select an appropriate distribution to describe the possible variation. Crystal Ball provides 16 standard distributions and a custom distribution to characterize the behavior or the assumption variable. Distribution selection should be based first on any available historical data. In the absence of historical data, one must rely on the underlying physics of the quantity to be described or select a simple distribution such as the triangular distribution and apply reasonable limits. Expert interviews and data gathering can provide critical information to assist in this process.

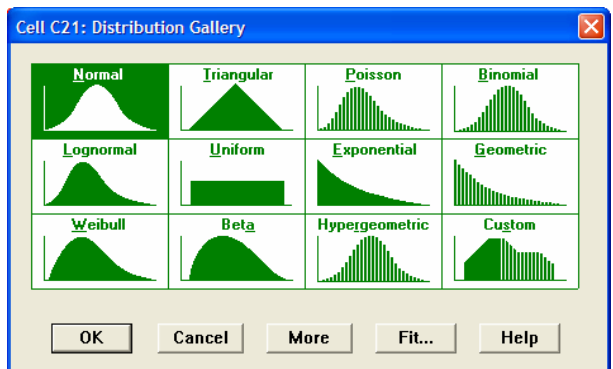

Figure 1: Crystal Ball Distribution Gallery

**2.2 The Monte Carlo Technique**

Monte Carlo simulation is a proven, efficient technique that only requires a random number table or a random number generator on a computer. A random number is a mathematically selected value that is generated to conform to a probability distribution. Monte Carlo simulation was named for Monte Carlo, Monaco, where the primary attractions are casinos containing games of chance, such as roulette wheels, dice, and slot machines, that exhibit random behavior.

The random behavior in games of chance is similar to how Monte Carlo simulation selects variable values at random to simulate a model. When you roll a die, you know that a 1, 2, 3, 4, 5, or 6 will come up, but you don't know which for any particular roll. It's the same with the variables that have a known range of values but an uncertain value for any particular time or event (e.g., interest rates, staffing needs, stock prices, inventory, phone calls per minute). The multiple scenarios created through simulation can be analyzed to give more insight into the risks and mechanisms of the spreadsheet model. When used correctly, Monte Carlo simulation can provide valuable insights not available through deterministic models.

**2.3 Interpretation of Results**

**The Simulation Process**

Spreadsheets provide a calculation engine for the model while Crystal Ball repeatedly samples from the input distributions. As the simulation progresses, key formulas identified by the user as forecasts are captured in memory for further analysis. For each of these forecast variables, there are two primary outputs from Crystal Ball, the forecast chart and the sensitivity chart. To understand how these charts can help identify potential faults in the model logic, it is necessary to first examine their relevance for risk quantification.

**Forecast Charts**

The forecast chart is a histogram that displays the range and frequency of different outcomes for an individual forecast. This information is used to calculate statistics about the probability or likelihood of different outcomes. For instance, one could calculate the probability of a financial loss occurring on a particular project where future revenue is uncertain. In addition to the graphical view, a full set of descriptive statistics is available for each simulation

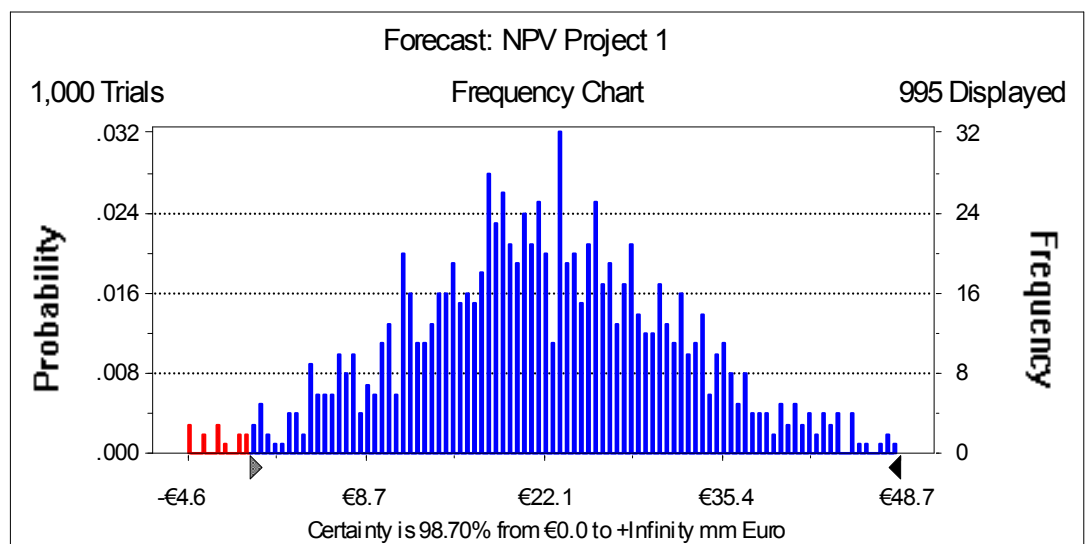

Figure 2: A typical forecast chart result

**Sensitivity Charts**

The second chart of importance is the sensitivity chart. Sensitivity charts display the relative relationship of an output forecast and each of the random assumption variables. This is represented by a correlation coefficient that measures the strength of the linear relationship between any two uncertain variables.

While the intent of sensitivity charts is to help identify key sources of variation, it is also an opportunity to perform a reality check. Seldom does a spreadsheet user attempt to model a problem without some knowledge of what factors are mission critical. In much the same way, most modelers have an intuitive understanding of the sort of results to expect from a particular forecast, making these two charts valuable in the identification of formula errors.

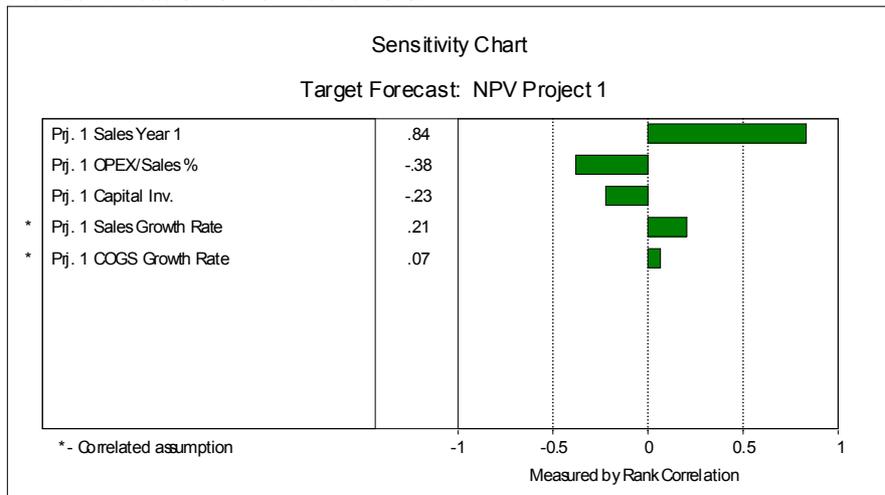

Figure 3: Typical sensitivity chart results

## 3 WARNING SIGNS

### 3.1 Errors During the Simulation Cycle

Before running the simulation, you can check for logic errors by clicking on the Single Step button on the tool bar (Figure 4). Each single step is a random trial in Crystal Ball. In one step, Crystal Ball generates a random number for each assumption, and Excel automatically recalculates the model. As the model recalculates for each trial, you may discover calculation errors due to unanticipated alternative scenarios.

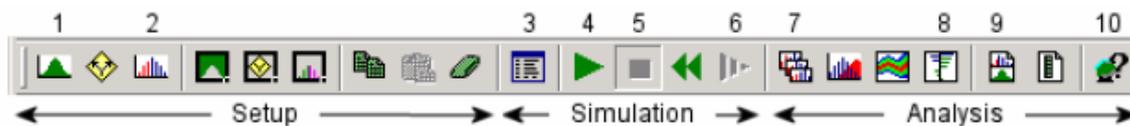

Figure 4: Tool Bar Buttons: (1) Define Assumption (2) Define Forecast (3) Run Preferences (4) Run (5) Stop (6) Single Step (7) Forecast Windows (8) Sensitivity Analysis (9) Create Report (10) Help

By default, a simulation stops whenever a calculation error occurs in a forecast cell. In this case a calculation error is defined as any formula which can not be resolved such as the square root of a negative number, division by zero, a bad look up table, or a financial function that does not converge due to a poor guess for an argument (for example IRR.) If this error occurs it identifies the location of the forecast cell and retains the underlying assumption values that lead to the calculation error.

Figure 5: Simulation with calculation error

Oftentimes, calculation errors are expected for certain financial calculations such as an Internal Rate of Return with a negative series of cash flows. In such cases, you can set the simulation to discard all trials that resulted in calculation errors. This adjustment, however, should only be done after careful consideration and ensuring that the calculation error is not due to some underlying flaw in the spreadsheet logic.

**3.2 Irregular Forecast Results**

**No Variation in Target Forecast**

One common cause of errors is the unintentional removal of formulas. Simulation add-ins often conflict with spreadsheet logic protection schemes. When cells are left unlocked, you are exposed to the risk of this innocent mistake. With complex logic spanning several sheets within the workbook, even experienced modelers can fall victim. Figure 6, which shows a forecast chart with a single value rather than a range of values, is a red flag that the logic chain may have been broken.

Figure 6: Forecast chart with incomplete model logic

**Out-of-Range Forecasts**

The forecast chart can also point to potential logical flaws by identifying out-of-range forecasts. While the root cause of the problem may be obscured, the results should be fairly obvious. The warning sign here is values that lie outside of theoretical limits. For example, some cash flow models must account for the fact that once a balance reaches zero, it must remain at zero for the remainder of the time horizon in the analysis. While the problem is relatively easy to fix from a formula standpoint, it is often overlooked because users seldom work with worst-case scenarios while building a deterministic model.

**3.4 Scenario Analysis**

The most effective way to understand forecasts outside the expected range is to find a set of assumption values that leads to the result in question. One method for accomplishing this is to apply the Scenario Analysis tool in Crystal Ball (Figure 7). First you specify a target range for the forecast, then Crystal Ball gathers all of the simulation trials that result in forecasts for that particular range. The tool also includes spreadsheet macros that paste the combination of inputs that led to a given output back into the original spreadsheet. This technique allows the user to investigate whether or not the forecast result is possible. On the other hand, if scenario analysis reveals a problem with the model logic, it is helpful to have a base case that replicates the error for auditing the formulas.

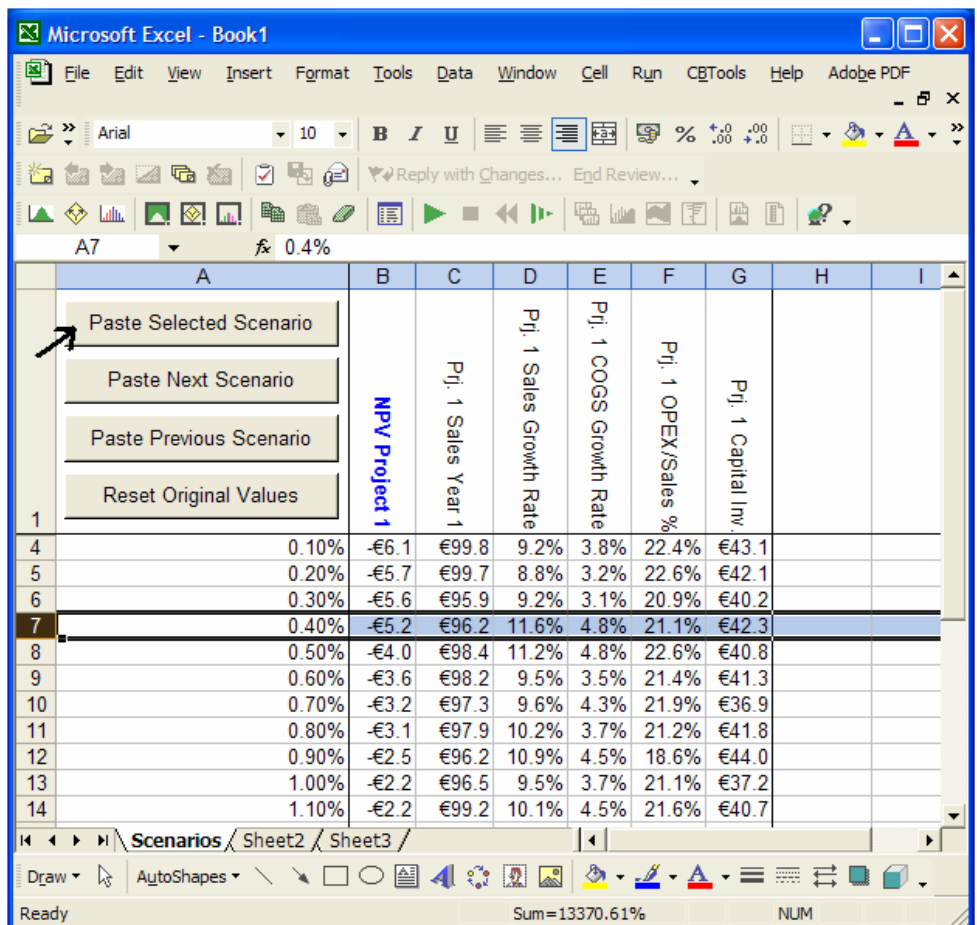

Figure 7: Scenario Analysis Output

**3.5 Sensitivity Sanity Check**

Results of the sensitivity chart should always be scrutinized to ensure that the results are consistent with the theoretical positive/negative relationship between an input assumption and the output forecast. Gut-feel instinct should not be ignored when the results of the sensitivity analysis conflict with prior experience. In Figure 8, the OPEX/Sales% appears to have the greatest effect on NPV, but experience suggests that Year 1 sales should have a much higher relative impact on the results of the simulation.

Identification of logical errors through Monte Carlo simulation: Hilary L. Emmett & Lawrence I. Goldman

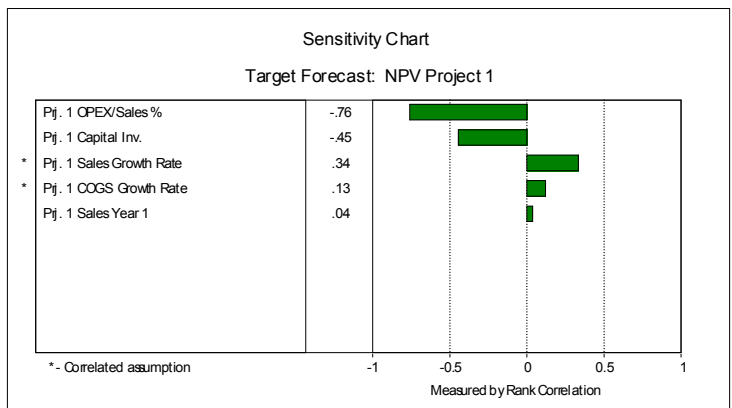

Figure 8: Inconsistent sensitivity chart results

If the results of the sensitivity chart are counter intuitive, two actions should be taken. First, review the parameters for input distributions. Inconsistent entry of percentage values is a common cause of such errors. Next, if discrepancies remain following a review of the input parameters, a review of the logic will be necessary. The tornado chart and Excel's auditing tools are useful for tracing formulas dependent on the input assumptions that appear to be under (or over) emphasized. In the case of the Year 1 sales, the discrepancy was traced to a formula that hard coded the Year 1 sales rather than referred to the input distribution (Figure 9).

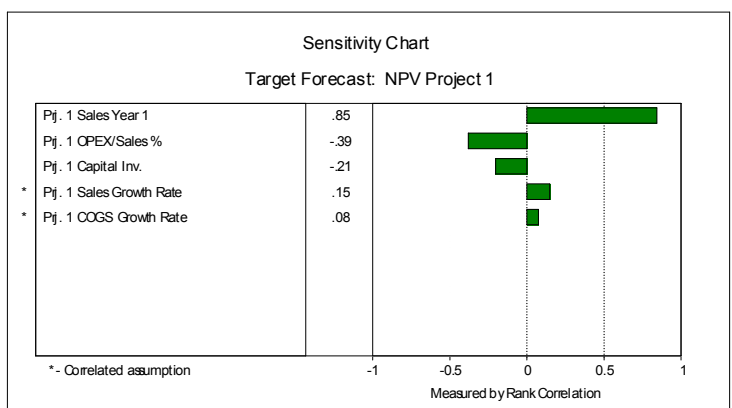

Figure 9: Counter intuitive sensitivity chart results

At first glance, the positive correlation between Net Present Value and Cost of Goods Sold (COGS) growth rate shown in Figure 9 could be cause for concern. If this were interpreted strictly as a cause and effect relationship, that would suggest that increasing costs leads to a direct increase in profits. A savvy modeler would consider the possibility of a formula error in the calculation of Net Sales in Row 12. Looking carefully at the sensitivity chart, you will notice that the assumptions for Sales Growth Rate and COGS Growth Rate are marked as correlated assumptions. This underlying relationship between the variables is important, as a strong positive correlation could mask the underlying cause and effect created by the model logic.

### 3.6 Resolving Discrepancies Using the Tornado Chart

Tornado charts provide an alternative sensitivity technique to rank correlation when trying to determine whether the negative relationship between COGS Growth Rate and NPV is caused by the underlying correlated assumptions or by a formula error. Rather than simultaneous sampling of all assumption variables, tornado charts isolate and change each variable separately (Figure 10).

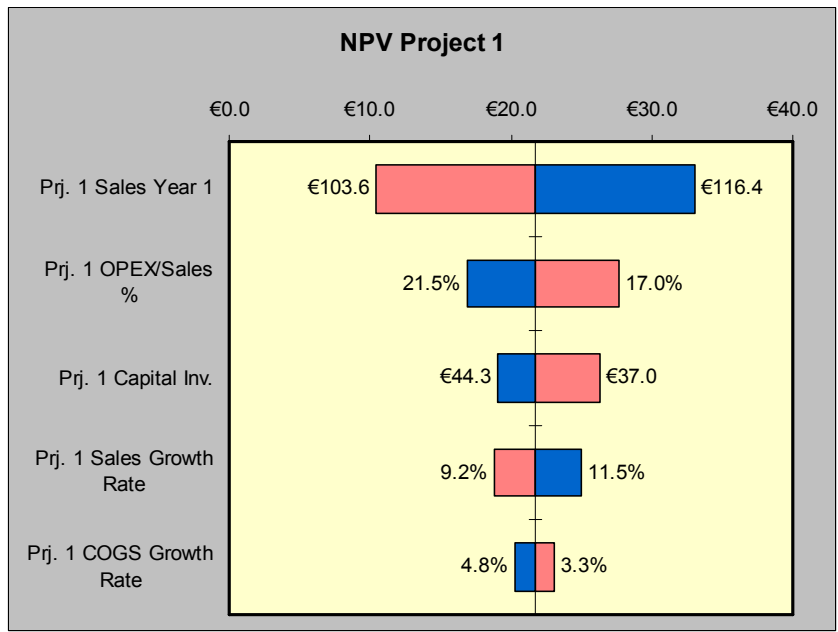

Figure 10: Tornado Chart results

By isolating the variables we are able to see the direct impact of change and understand the underlying logic. In this example, we do not see any cause for doubting the model, an increase in COGS Growth rate causes a small decrease in Project NPV.

**3.7 Self-fulfilling Prophecies**

One must always be weary of using a gut-feel based approach to spreadsheet audits. Familiarity with the spreadsheet makes the developer a less than objective reviewer. In Ayalew Yirsaw's Interval Based approach, this problem is addressed by the spreadsheet developer documenting expected intervals for output results to serve as a guide for an independent auditor. Another way to avoid such problems is to use a technique referred to as back-casting. Monte Carlo simulation can sample random historical data points to test the model performance with real data.

**4 CONCLUSION**

Performing simulation analysis for future outcomes can help spreadsheet modelers to reduce their risks and dramatically improve the quality of their decisions. By moving to a probabilistic approach to spreadsheet forecasting, analysts can better quantify the risks inherent in their models and gain insights not available through traditional deterministic approaches. The stress placed on the model in the simulation process exposes weaknesses in the formulas and provides clues for tracing breaks in the logical flow.